# Prediction of Sea Level Rise near Shanghai

Yi ZHENG

Business School, Shanghai Jian Qiao University, Shanghai 201306，China.

*Abstract*—**Firstly, by establishing a prediction model for global sea-level rise and calculating with Maple, it is showed: the global sea level rise rate in 2009 is 2.68mm/a. The height and rate of global sea-level rise will be about 9.11cm and 3.22mm/a in 2020,. Based on the study and the actual land subsidence in Shanghai Lingang New City, the rate of relative sea-level rise near Lingang New City is calculated to be 12.68mm/a in 2009. Then, through setting of the extrapolation prediction model with linear trend term and a significant cycle of tidal, the rise rate of average sea-level near Lingang New City was predicted. The result showed: it will be 0.33mm/a in 2020.**

## I. Introduction

Sea-level rise is a slow, long and beyond retrievable ocean disaster. It often aggravates storm tide. It is often further aggravated the ocean disasters, such as storm tide, red tide, saltwater intrusion and salinization. The threat will be a long time. Especially, sea-level rise has accelerated the trend in recent years. Sea-level rise has led to a lot of coastal big cities like London, New York and Shanghai was forced to spend billions of dollars for preventing these disasters.

It is generally believed that sea-level rise is mainly due to the Earth's climate warming. The specific reasons include: firstly, the thermal expansion by volume of sea water; secondly, the sea water imported from lakes, groundwater, mountain glaciers is increased due to global warming; at last, it is the results of accelerated melting of the Antarctic and Greenland ice sheets [1].

In addition to the factor of global warming, to the long time scale, sea-level change also is affected by changes in Earth's structure, activities of great ice age and the atmosphere, ocean itself and other relevant factors. For the regional or local changes in sea-level, there is a close relationship with the global mean sea-level. In the same time, they are also related to the structure, weather and ocean conditions of their own different geographical areas or different parts.

Along the Chinese coast, in the past 30 years, sea-level rose by an average rate of 2.6mm/a, higher than the global 1.8mm/a in average[2]. A report issued by the State Oceanic Administration of China further shows that the rise height of sea-level of East China Sea will be 37mm in the next 10 years, higher than the Bohai Sea's 29mm, Yellow Sea's 31mm and the South China Sea's 30mm[3]. So the rise of sea-level will greatly affect the building for Shanghai Lingang New City, where is located in the edge of East China Sea and is planned as a modern integrated seaside city by the government.

So it will be helpful for Lingang New City about its reasonable city layout and marine development to research the forecast of the rising of sea-level. So far, there are a few studies for the future change of Chinese coastal sea-level, but in this study, it is few for the prediction to Lingang New City and their prediction accuracy is limited because the methods used are simple linear model. To make up the lack of research on sea-level rise for Lingang New City and to improve forecast accuracy，a cycle extrapolation prediction model was set up in this paper.

## II. Prediction for the Rise Rate of Relative Sea Level

For providing reliable guidance in decision making processes, accurate and robust to distribution shifts are often to be required[4].To research on the sea-level rise, the assumptions are as follows:

- The thermal expansion of sea water caused by temperature is the main factor to affect sea-level change. So temperature changes significantly affect sea-level and change of sea surface temperature would lead to sea-level changes.

- Sea-level change in specific areas can be integrated by the region's some significant cycled tidal.

In this paper, the relative sea-level means that the actual sea-level change of an area is affected by the local land vertical movement---the slow crustal movements and the local land subsidence. The relative sea-level change in a region is global sea-level rise add the local land rise or down values. The average sea-level is the ideal calm sea where the water level is equal to the average values of observations. The concepts about average sea-level are different if the time of observation is different, such as daily average sea-level, monthly average sea-level, the yearly average sea-level and the average sea-level for many years and so on.

## A. The model for global sea-level based on the concentration of $CO_2$

Increasing of Global temperature have lead to sea-levels rising. Comparing the global atmospheric temperature and concentration of $CO_2$ in the last 400,000 years, it is found that their overall trend was almost the same[2]. So it is reasonable to use the model based on the concentration of $CO_2$ for forecasting sea-level in this paper.

In 1985 Wigley set up a sea-level forecast equation which linked the temperature change, the greenhouse affect and ocean thermal diffusion well together[5]. He put forward that the emission concentration of $CO_2$ may be estimated by formula (1):

$$C(t) = C_0 \exp[Bt \exp(\alpha t)] \quad (1)$$

Where $B = 5.59 \times 10^{-4}; \alpha = 8.69 \times 10^{-3}; C_0 = 270.0$, and t is the ordinal year since 1850.

The amount of radioactive force under Greenhouse effect can be written as a function of emission concentration of $CO_2$

$$F(t) = a_0 + b \ln[C(t)/270.0] \quad (2)$$

Where $a_0 = 1.81 W/m^2$; $b = 2.95 W/m^2$.

Temperature change caused by the increase in concentration of $CO_2$ emission was looked as a function of time t:

$$T(t) = \beta t \ln[\gamma t C(t)/270.0] - 0.5 \quad (3)$$

Where $\beta = 0.00593$, $\gamma = 0.0114$.

$$Z(t) = [4.13 + 2.6 F(t)] T(t) \kappa^{0.221} \quad (4)$$

In formula (4), Z(t) expresses rising or falls values of sea-level in a certain time period. $\kappa$ is the ocean thermal expansion coefficient, its range is from 0.5 to 3.0 ($cm^2/s$). Based on the calculated result from formula (1), substituted F(t) and T(t) to Z(t) in the formula (4) with the corresponding values which calculated from formulas (2) and (3), the change values of sea-level will be available.

## B. Improvement and application of the model for global sea-level

In the relevant research literatures on the sea-level rise, the value of the ocean thermal diffusion coefficient $\kappa$ in the above model is always limited to a range, such as: Zhang jin-wen got the value $\kappa$ in turn In an interval in his research, but a relatively fixed and the appropriate value of $\kappa$ has not been given[3]. It is inconvenient for the research of sea-level rise.

For ascertain the value of the ocean thermal diffusion coefficient, the global data of $CO_2$ emissions concentration has been collected from 1870 to 2006 in this paper (Tab.1). In the other hand, some scientists reported that the earth's average temperature increased 0.19 degrees Celsius every 10 years in the past 25 years, and will be rise up to 7 Celsius in this century. According to the analysis and calculation by these data, this paper believes: it is more appropriate when the value of the oceans thermal diffusion coefficient is 3.0.

Tab.1 The global data of $CO_2$ emissions concentration from 1870 to 2006

| Year | 1870 | 1880 | 1890 | 1900 | 1910 | 1920 |
|---|---|---|---|---|---|---|
| $CO_2$ emissions concentration (ppm) | 290 | 291 | 292 | 293 | 294 | 295 |
| Year | 1930 | 1940 | 1950 | 1960 | 1970 | 1980 |
| $CO_2$ emissions concentration (ppm) | 300 | 305 | 310 | 320 | 328 | 340 |
| Year | 1990 | 2000 | 2004 | 2005 | 2006 | |
| $CO_2$ emissions concentration (ppm) | 353 | 370 | 378.4 | 379.19 | 381.2 | |

For convenience of calculation, the temperature and sea-level in 1980 is used as a standard, their values were assumed equal to zero. By taking different t and $\kappa = 3.0$, substituting formula (1) into formula (2) and (3), then (2) and (3) formulas were substituted in formula (4) again, so the incremental value of global sea-level rise to the year of 1980 was available. If t was got to be 170, 180, 190 and 200 separately, some prediction values of future global sea-level rise were calculated from the above formulas. (listed in Table 2)

Tab.2 The prediction value of global sea-level rise

| Year $\kappa = 3.0$ | 2020 | 2030 | 2040 | 2050 |
|---|---|---|---|---|
| Height of sea-level rise (cm) | 9.11 | 12.6 | 16.81 | 35.43 |

By the table 2, the global sea-level in 2050 will be 35.43cm higher than sea-level in 1980. it is very dangerous. Derivation of t on formula (4) can get the prediction equation of global sea-level rise rate. Take t = 159, the Estimate of global sea-level rise rate in 2009 is 2.68mm /a. in the other hand, the value of actual average sea-level rise rate in 2008 has reached 20mm / a. It shows: the improved simulation in this paper has a good accuracy. Similar, if taking t = 180, the simulation can predict the global sea-level rise rate in 2020 to be 3.2mm/a.

Land subsidence has great influence on the relative sea-level in Shanghai. From 2001 to 2006, the annual average land subsidence of Lingang New City in Shanghai was 5mm to 15mm[6]. Taking the median of the range, its average rate of land subsidence is assumed to be 10mm/a. So by the definition of relative sea-level, the rise rate of relative sea-level near Lingang New City is equal the rate of global sea-level rise add its average rate of land subsidence, i.e. 2.68 +10 = 12.68 (mm/a).

## III. PREDICTION FOR THE RISE RATE OF SEA LEVEL BY THE MODEL OF EXTRAPOLATION CYCLE METHOD

The above model got local relative sea-level rise by measuring the rate of global sea-level rise and combined with local average rate of land subsidence, but it can't measure the absolute sea-level changes. Because the sea-level changes in itself periodicity and trend, the harmonic constants of the periodic vibration vary by different region, and their selection is significant for the predicting of monthly average water level, so the further model was established in the next text to predict more accurately the absolute sea-level changes near Lingang New City.

Changes of sea-level included a linear trend and a number of periodic vibrations. The trend reflects the long-term sea-level rise. When the changes in the trend were written a linear function of time, the periodic vibration is divided into several sub-tidal, the analysis equation for trend and periodic changes in sea-level is follows[6]:

$$H(t) = a + bt + \sum_{k=1}^{n} A_k \sin(2\pi t / T_k + \phi_k) + \varepsilon_k \quad \cdots(5)$$

Where a, b and $\varepsilon_k$ are constant, linear and random error, $T_k$, $A_k$ and $\Phi_k$ were the signal cycles, amplitude and phase, n is the total number of long-period tidal.

The purpose of harmonic analysis is to extract the periodic components of the time series, that is, the amplitude, period and phase of periodic signal will be exacted by this method. The harmonic analysis for the rise of sea-level H(t) is to determine $T_k$, $A_k$ and $\Phi_k$. If $T_k$, is known, $A_k$ and $\Phi_k$, even a and b can be estimated by the least square approximation.

The frequencies and their amplitudes of a sequence can be analyzed by Fast Fourier Transform. The 16-year data of Luchaogang (observed by the Nanhui mouth station near Lingang New City) was analyzed by fast Fourier transform [7-10]. It showed: the amplitude is 0.3597cm to MN for 18.61 years, its frequency is $3.79 \times 10^{-3}$ / month, corresponding to the cycle of 22 years. In addition, the analysis also got a sub-tide, which has a cycle of 1.57 years, a frequency of 0.053 / month, a amplitude of 0.2101cm; and a Sa, which has a amplitude of 0.12cm, a frequency of 0.845 / month, a cycle of 1.18 years; and a Saa, which has a amplitude of 0.096cm, a frequency of 0.5/month, a cycle of 0.5 years. Put these results into formula (5), it is:

$$H(t) = a + bt + 0.3597\sin(0.00378789t + f_1) + 0.2101\sin(0.053t + f_2) + 0.12\sin(0.845t + f_3) + 0.096\sin(0.5t + f_4) \quad (6)$$

Based on these results and existing data, got MN of 18.61 years, sub-tide, Sa and Saa of 1.57 years, formula (6) was estimated with a non-linear regression. The results are: a= 259.2708, b= 0.0382, $f_1$= 491.3048, $f_2$= -1.9075, $f_3$= 13.9185, $f_4$= -110.9758,

Then formula (6) become into:

$$H(t) = 259.2708 + 0.0382t + 0.3597\sin(0.00378789t + 491.3048) + 0.2101\sin(0.053t-1.9075) + 0.12\sin(0.845t + 13.9185) + 0.096\sin(0.5t-110.9758) \quad (7)$$

where t = 1 is represented in January 1987, as the first item of the month sequence. So the average sea-level for twelve months in 2020 can be predicted if t is valued from 408 to 419. (see Table 3)

Tab.3 monthly mean sea-level in 2020

| Month | 1 | 2 | 3 | 4 |
|---|---|---|---|---|
| Average sea-level /cm | 273.90 | 274.09 | 274.26 | 274.34 |
| Month | 5 | 6 | 7 | 8 |
| Average sea-level /cm | 274.34 | 274.29 | 274.26 | 274.29 |
| Month | 9 | 10 | 11 | 12 |
| Average sea-level /cm | 274.38 | 274.48 | 274.53 | 274.52 |

Derivation of t on formula (7), then getting t from 408 to 419, 12 months average monthly rate of sea-level rise in 2020 is available. Taking the average, it shows: average rate of sea-level rise in Lingang New City will be 0.033cm/ a in 2020.

## IV. CONCLUSIONS AND DISCUSSION

In this paper, the global sea-level rise rate in 2009 is calculated to be 2.68mm /a by the model based on the global $CO_2$ concentration. Use the above result, the rise rate of relative sea-level near Lingang New City is estimated to be 12.68 (mm/a). With sea-level extrapolation prediction model by cycle method, the rise rate of monthly average sea-level near Lingang New City is 0.33mm/a in 2020.

Because the sea-level data obtained is short and incomplete, and the tidal data is very random so that the noise of the data may be large for small sea-level changes, they may affect the accuracy of forecasts. But the research results still have a reference value for the prevention and mitigation of disaster and urban planning to Shanghai Lingang New City. In particular, The model used in this paper has a certain value in the practice of sea-level forecasting.